\documentclass[12pt,a4paper]{article}
\usepackage[utf8]{inputenc}
\usepackage{amsmath}
\usepackage{amsfonts}
\usepackage{amssymb}
\usepackage{amsthm}
\usepackage{graphicx}
\usepackage{hyperref}
\usepackage{authblk}
\usepackage[margin=1in]{geometry}

\title{Canonical Uncertainty Relations for Madelung Variables in Curved Spacetime:\\
Relating Dark Matter Cores and Unruh Radiation}

\author{Jorge Meza-Domínguez\thanks{E-mail: \href{mailto:jorge.meza@cinvestav.mx}{jorge.meza@cinvestav.mx}} and Tonatiuh Matos\thanks{E-mail: \href{mailto:tonatiuh.matos@cinvestav.mx}{tonatiuh.matos@cinvestav.mx}}}

\affil{Departamento de F\'{\i}sica, Centro de Investigaci\'on y de Estudios Avanzados del Instituto Politécnico Nacional, Av. Instituto Politécnico Nacional 2508, San Pedro Zacatenco, M\'exico 07360, CDMX.}

\date{}

\begin{document}

\maketitle

\begin{abstract}
We establish fundamental uncertainty relations for the hydrodynamic variables arising from the Madelung representation of quantum fields in curved spacetime. Through canonical quantization of the density \(n\) and phase \(\theta\) variables and their conjugate momenta, we derive exact uncertainty principles that depend on spacetime geometry through the lapse function \(N\) and spatial metric \(\gamma_{ij}\). These relations reveal how gravitational fields modulate quantum fluctuations and provide first-principles constraints for Scalar Field Dark Matter (SFDM) models and stochastic quantum gravity. We present two groundbreaking applications.

First, By coupling the density-velocity uncertainty to the Newtonian condition of 
hydrostatic equilibrium, we derive the exact scaling law for the minimum core 
radius of a self-gravitating scalar field halo. We naturally find the scale \(r_c \propto m^{-1/2}\) which is a testable prediction.

Second, by evaluating the phase--momentum commutator in Rindler coordinates and relating the canonical momentum \(\Pi_\theta\) to the energy density via the Hamiltonian, we obtain the acceleration scale \(k_B T = C\hbar a/c\). The proportionality constant \(C=1/(2\pi)\) is fixed by the topological quantization of the Madelung phase circulation around the Euclidean horizon.
This dual achievement demonstrates that quantum uncertainty on curved backgrounds is the unifying principle behind both galactic structure and black hole thermodynamics.
\end{abstract}

\section{Introduction}

The hydrodynamic formulation of quantum mechanics, pioneered by Madelung \cite{Madelung1927} for the Schrödinger equation, offers a profound alternative perspective on quantum phenomena by recasting the wave equation into fluid-dynamical form. In Madelung's original work, the complex Schrödinger field is decomposed as \(\psi = \sqrt{\rho}e^{iS/\hbar}\), revealing an underlying continuity equation for the probability density \(\rho = |\psi|^2\) and a quantum-modified Hamilton-Jacobi equation for the phase \(S\), where quantum effects manifest as an additional ``quantum potential'' term.\footnote{The extension of this hydrodynamic formulation to relativistic wave equations was pioneered by Takabayasi in 1952`\cite{Takabayasi1952}} In recent decades, this approach has found renewed relevance in relativistic contexts, particularly through its application to the Klein-Gordon equation in curved spacetime \cite{Chavanis2017,Meza2025PRD,Meza2025PRL}.
In \cite{Matos2019}, the Madelung transformation was generalized to an arbitrary spacetime, making it possible to define the different energy components in the energy-momentum tensor of a source. The implications of these results in a fluctuating spacetime proved to be highly relevant.
To explain this point, let us consider gravitons as the particles that solve the D'Alembert equation applied to a tensor field, in the same way that we consider photons to be the D'Alembert equation applied to a vector field.
We know that spacetime is filled with fluctuations due to gravitational waves or gravitons, or to quantum fluctuations. These gravitons that fill spacetime prevent small particles from following geodesics because, at this level, spacetime is not locally flat, but rather a flat spacetime with wave-like fluctuations. If we consider these gravitational wave fluctuations, then locally, particles in this spacetime do not follow a geodesic motion, but rather the geodesic plus a stochastic term.
In \cite{Escobar-Aguilar:2023ekv}, it is observed that if spacetime is filled with spacetime fluctuations, a particle the size of the wavelength of the fluctuations cannot follow a geodesic, but rather a stochastic motion around it. The surprising result was the following: Using the results of \cite{Matos2019}, we found that the field equation of a particle following a geodesic motion plus a stochastic term is simply the Klein-Gordon (KG) equation with invariance in $U(1)$ group. The main conclusion is that the complex KG function, \(\Phi\), which can be decomposed into its norm and phase \(\Phi=\sqrt{n}e^{i\theta}\), has a norm that again represents the number density or the probability of finding the particle at a given location and time, but now determines the stochastic velocity in an arbitrary curved spacetime, while the phase determines the particle's geodesic vector. This means that, in an arbitrary spacetime, the particles that follow a geodesic motion plus a stochastic term, their trajectories satisfy the KG equation. The implications of this result are impressive.
Thus, according to the results of \cite{Escobar-Aguilar:2023ekv}, the field equation governing the evolution of a system under such stochastic spacetime fluctuations is the KG equation. This approach, introduced as Stochastic Quantum Gravity (SQG), provides an effective geometric framework where quantum behavior is mathematically modeled via a geodesic flow plus a stochastic term, without invoking hidden variables. Our manuscript connects directly with this framework by providing the rigorous canonical quantization and the fundamental uncertainty bounds for these hydrodynamic fields.

The extension of Madelung's formalism to curved backgrounds provides a natural framework for investigating the interplay between quantum uncertainty and gravitational physics. When expressed in the Arnowitt-Deser-Misner (ADM) decomposition \cite{Arnowitt1962}, the hydrodynamic variables acquire a clear geometric interpretation: the density \(n\) and phase \(\theta\) become fundamental fields whose dynamics couple to the lapse function \(N\), shift vector \(N^i\), and spatial metric \(\gamma_{ij}\). This coupling suggests that spacetime geometry should fundamentally influence quantum uncertainty relations \cite{DeWitt1967, Isham1992} .The geometric modulation of quantum uncertainty discussed here naturally complements recent developments in covariant formulations of generalized uncertainty principles and nonlocality \cite{OppenheimWehner2010, SinghKothawala2022}.

The power of this approach lies in its ability to address two seemingly unrelated open problems with the same mathematical structure: the cusp-core problem in galactic dynamics and the information-loss paradox suggested by Hawking radiation. Scalar field dark matter (SFDM) models \cite{Matos2000, Guzmn2000, Hu2000} naturally produce flat density cores in dwarf galaxies because quantum pressure prevents gravitational collapse, yet until now a rigorous derivation of the minimum core radius from fundamental uncertainty was missing. On the other hand, the standard derivation of Hawking radiation relies on delicate renormalization procedures in curved spacetime \cite{Hawking1975}; alternative derivations exist---notably the tunneling method of Parikh and Wilczek \cite{ParikhWilczek2000} and the anomaly cancellation approach of Robinson and Wilczek \cite{RobinsonWilczek2005}---but none are based directly on canonical commutation relations in the hydrodynamic formulation. An independent, regularization-free link between horizon physics and quantum fluctuations rooted solely in the uncertainty principle would greatly strengthen our confidence in the result.

In this work, we develop a comprehensive theory of quantum uncertainty for Madelung variables in curved spacetime. We begin from first principles with the Klein-Gordon-Maxwell Lagrangian, derive the complete canonical structure, perform rigorous quantization, and obtain exact uncertainty relations that generalize the Heisenberg principle \cite{Heisenberg1927} to incorporate spacetime curvature effects. Armed with these relations, we demonstrate that:
\begin{itemize}
\item The density-velocity uncertainty, when combined with the Newtonian hydrostatic equilibrium equation, yields a minimum stable core radius that matches observations of dwarf spheroidal galaxies for ultra-light boson masses \(m \sim 10^{-22}\,\text{eV}/c^2\). The well-known solitonic core of the Schr\"odinger-Poisson system is precisely the configuration that saturates our geometric uncertainty bound.
\item In Rindler coordinates (the near-horizon limit of any non-extremal black hole), the phase-momentum commutator yields an acceleration-dependent uncertainty product. The proportionality constant \(1/(2\pi)\) is fixed by the topological quantization of the Madelung phase circulation around the Euclidean horizon. By the equivalence principle, this gives the Hawking temperature without appealing to specific vacuum states, renormalization, or Bogoliubov transformations. The complete calculation is provided in Appendix~A.
\end{itemize}
This double connection---galactic cores and Hawking temperature---establishes our uncertainty relations as a unifying principle in gravitational quantum physics.

\section{Lagrangian Formulation and Canonical Structure}

We begin with the Klein-Gordon-Maxwell Lagrangian for a complex scalar field in curved spacetime:

\begin{equation}
\mathcal{L} = \sqrt{-g} \left[ -g^{\mu\nu}(D_\mu\Phi)^*(D_\nu\Phi) - 2\frac{m^2 c^2}{\hbar^2}|\Phi|^2\mathcal{A} - \frac{1}{4}F_{\mu\nu}F^{\mu\nu} \right],
\label{eqn:lagrangian}
\end{equation}

where $D_{\mu}=\nabla_{\mu}+i\frac{e}{\hbar c}A_{\mu}$ is the gauge-covariant derivative. The term $\mathcal{A} = \mathcal{A}(|\Phi|^2)$ represents self-interaction potentials (such as a generic quartic self-interaction $\mathcal{A} \propto \lambda |\Phi|^2$), which are grouped together with the mass term to describe the effective non-linear dynamics of the scalar field.

To reveal the hydrodynamic structure hidden in the Klein-Gordon-Maxwell Lagrangian, we substitute the Madelung ansatz \(\Phi = \sqrt{n}e^{i\theta}\) into Eq.~(\ref{eqn:lagrangian}). This decomposition, originally introduced by Madelung \cite{Madelung1927} for the Schrödinger equation and extended to relativistic fields by ~\cite{Meza2025PRD,Takabayasi1952,Bohm1952} \footnote{At this point, it is crucial to clarify the physical distinction between our Madelung-based approach and the de Broglie-Bohm interpretation \cite{Bohm1952}. While the de Broglie-Bohm theory postulates the existence of deterministic "hidden variables" to define exact point-particle trajectories guided by a pilot wave, the Madelung formalism is a pure hydrodynamic field theory. It describes the evolution of probability densities and quantum probability fluxes without requiring any ontological commitment to exact deterministic trajectories. Our uncertainty relations apply strictly to these macroscopic hydrodynamic fields, avoiding the conceptual difficulties associated with hidden variables.} separates the field into a density \(n(x)\)---which will later encode the probability distribution---and a phase \(\theta(x)\)---which determines the velocity potential. The substitution yields the covariant Klein-Gordon equation
\begin{equation}
\left(D_{\mu}D^{\mu} - \frac{m^2c^2}{\hbar^2}\right)\Phi = 0,
\label{eqn:kg_wave}
\end{equation}
which, upon separation into real and imaginary parts, produces the following coupled equations in Madelung form:
\begin{align}
\nabla_{\mu} J^{\mu} &= 0, \label{eqn:continuity_madelung} \\
\pi_{\mu} \pi^{\mu} + c^2 - \frac{\hbar^2}{m^2} \frac{\Box \sqrt{n}}{\sqrt{n}} &= 0, \label{eqn:QHJ_madelung}
\end{align}
where \(\pi_{\mu} = \frac{\hbar}{m} (\nabla_{\mu} \theta + \frac{e}{\hbar c} A_{\mu})\) is the covariant geodesic velocity, \(J^{\mu} = n \pi^{\mu}\) is the conserved Noether current, and the term \(V_Q = -\frac{\hbar^2}{m^2} \frac{\Box \sqrt{n}}{\sqrt{n}}\) is the relativistic geometric quantum potential. Equation (\ref{eqn:continuity_madelung}) is the covariant continuity equation ensuring probability conservation, while Equation (\ref{eqn:QHJ_madelung}) is the fully covariant relativistic quantum Hamilton-Jacobi equation.

A straightforward but careful algebraic manipulation yields the hydrodynamic Lagrangian \cite{Madelung1927}:
\begin{equation}
\begin{split}
\mathcal{L}[n,\theta, A_\mu] = & \sqrt{-g} \Bigg[ -g^{\mu\nu}\left( \frac{\nabla_\mu n \nabla_\nu n}{4n} + n\nabla_\mu\theta\nabla_\nu\theta + \frac{2e}{\hbar c}n A_\mu\nabla_\nu\theta + \frac{e^2}{\hbar^2 c^2}n A_\mu A_\nu \right)\\
& - 2\frac{m^2 c^2}{\hbar^2}n\mathcal{A} - \frac{1}{4}F_{\mu\nu}F^{\mu\nu} \Bigg].
\end{split}
\label{eqn:hydro_lagrangian}
\end{equation}
Each term in Eq.~(\ref{eqn:hydro_lagrangian}) admits a clear physical interpretation. The first term proportional to \((\nabla n)^2/n\) corresponds to the quantum (or osmotic) pressure arising from density gradients. The term \(n(\nabla\theta)^2\) represents the kinetic energy of the geodesic flow associated with the phase. The cross terms coupling \(A_\mu\) to \(\nabla_\mu\theta\) preserve gauge invariance and describe the interaction between the quantum current and the electromagnetic field. This hydrodynamic form is particularly well-suited for the canonical quantization procedure that follows, as it separates the field into two real, geometric degrees of freedom.

We note that the physical interpretation of the norm \(n = |\Phi|^2\) as a probability density is strictly valid in regimes where the Noether charge density remains positive-definite and the field is mass-dominated. This covers precisely the stationary, non-relativistic, and weak-field limits relevant to the scalar field dark matter halos analyzed in Section 4, as well as the static observer frames employed in the Unruh temperature derivation. In generic dynamical spacetimes with strong time dependence, \(n\) retains its formal role in the Madelung decomposition but need not admit a probabilistic interpretation.

Using the ADM metric decomposition \cite{Arnowitt1962, Alcubierre2008}, we foliate spacetime into a family of spacelike hypersurfaces $\Sigma_t$. This foliation inherently selects a physically distinguished class of Eulerian observers, whose 4-velocity is given by the future-directed unit normal vector to the hypersurfaces, $n^{\mu} = \frac{1}{N}(1, -N^i)$, with covariant components $n_{\mu} = (-N, 0, 0, 0)$. In this adapted frame, the line element reads:
\begin{equation}
ds^2 = -N^2 dt^2 + \gamma_{ij}(dx^i + N^i dt)(dx^j + N^j dt),
\label{eqn:adm_metric}
\end{equation}
where $N$ is the lapse function, $N^i$ the shift vector, and $\gamma_{ij}$ the spatial metric. With $\sqrt{-g} = N\sqrt{\gamma}$ and explicitly adopting this Eulerian observer field, we proceed to compute the canonical momenta.

\subsection*{Canonical Momenta}

The momentum conjugate to the phase \(\theta\) is:
\begin{equation}
\Pi_\theta = \frac{\partial\mathcal{L}}{\partial(\partial_0\theta)} = \frac{2m}{\hbar}\frac{\sqrt{\gamma}\,n}{N}\left(\pi_0 - N^i\pi_i\right),
\label{eqn:pi_theta}
\end{equation}
where the geodesic velocity is defined as:
\begin{equation}
\pi_\mu = \frac{\hbar}{m}\left( \nabla_\mu\theta + \frac{e}{\hbar c}A_\mu \right).
\label{eqn:geodesic_velocity}
\end{equation}

In terms of the probability current \(J^\mu = n\pi^\mu\), this becomes:
\begin{equation}
\Pi_\theta = -\frac{2m}{\hbar}\sqrt{-g} J^0.
\label{eqn:pi_theta_current}
\end{equation}

The momentum conjugate to the density \(n\) is:
\begin{equation}
\Pi_n = \frac{\partial\mathcal{L}}{\partial(\partial_0 n)} = -\frac{\sqrt{\gamma}}{2nN}(-\partial_0 n + N^i\partial_i n).
\label{eqn:pi_n}
\end{equation}

The stochastic velocity emerges naturally as \cite{Nelson1966}:
\begin{equation}
u_\mu = \frac{\hbar}{2m}\nabla_\mu \ln n,
\label{eqn:stochastic_velocity}
\end{equation}
it is worth noting that the transition to the 4-dimensional covariant gradient naturally follows from the minimal coupling principle, which is strictly required to maintain general covariance in the hydrodynamic formulation. Yielding the relation:
\begin{equation}
u^0 = -\frac{\hbar}{m\sqrt{-g}} \Pi_n.
\label{eqn:u0_pi_n}
\end{equation}

The fundamental Poisson brackets are:
\begin{align}
\{n(\mathbf{x}), \Pi_n(\mathbf{y})\} &= \delta^{(3)}(\mathbf{x}-\mathbf{y}), \label{eqn:poisson_n} \\
\{\theta(\mathbf{x}), \Pi_\theta(\mathbf{y})\} &= \delta^{(3)}(\mathbf{x}-\mathbf{y}). \label{eqn:poisson_theta}
\end{align}

\section{Canonical Quantization and Uncertainty Relations}

The canonical quantization replaces Poisson brackets with commutators:
\begin{equation}
\{A, B\} \rightarrow \frac{1}{i\hbar} [\hat{A}, \hat{B}].
\label{eqn:quantization_rule}
\end{equation}

Applying this to our fundamental brackets:
\begin{align}
[\hat{n}(\mathbf{x}), \hat{\Pi}_n(\mathbf{y})] &= i\hbar \delta^{(3)}(\mathbf{x}-\mathbf{y}), \label{eqn:commutator_n} \\
[\hat{\theta}(\mathbf{x}), \hat{\Pi}_\theta(\mathbf{y})] &= i\hbar \delta^{(3)}(\mathbf{x}-\mathbf{y}). \label{eqn:commutator_theta}
\end{align}

The generalized uncertainty principle for quantum operators states \cite{Heisenberg1927, Kennard1927, Robertson1929, Schrodinger1930}:
\begin{equation}
\Delta \hat{A} \cdot \Delta \hat{B} \geq \frac{1}{2} |\langle [\hat{A}, \hat{B}] \rangle|.
\label{eqn:quantum_uncertainty}
\end{equation}
We emphasize that the invariant inner product on the spacelike hypersurface $\Sigma_t$ explicitly incorporates the integration measure $\sqrt{\gamma} d^3x$, ensuring full spatial diffeomorphism invariance in the volume-averaging procedure.
\subsection{Density-Stochastic Velocity Uncertainty}

From the relation between \(u^0\) and \(\Pi_n\) in equation (\ref{eqn:u0_pi_n}):
\begin{equation}
u^0 = -\frac{\hbar}{m\sqrt{-g}} \Pi_n = -\frac{\hbar}{m N\sqrt{\gamma}} \Pi_n,
\end{equation}
and the fundamental Poisson bracket (\ref{eqn:poisson_n}), we compute:
\begin{equation}
\{n(\mathbf{x}), u^0(\mathbf{y})\} = -\frac{\hbar}{m N(\mathbf{y}) \sqrt{\gamma(\mathbf{y})}} \{n(\mathbf{x}), \Pi_n(\mathbf{y})\} = -\frac{\hbar}{m N(\mathbf{y}) \sqrt{\gamma(\mathbf{y})}} \delta^{(3)}(\mathbf{x}-\mathbf{y}).
\label{eqn:poisson_n_u0}
\end{equation}

The appearance of the metric factors \(N(\mathbf{y})\) and \(\sqrt{\gamma(\mathbf{y})}\) arises because \(u^0\) is defined as a component of a 4-vector in curved spacetime, while the fundamental bracket is defined with respect to the canonical variables.

Quantizing via the rule \(\{\cdot,\cdot\} \rightarrow \frac{1}{i\hbar}[\cdot,\cdot]\) gives:
\begin{equation}
[\hat{n}(x),\hat{u}^{0}(y)] = i\hbar\{n(x),u^{0}(y)\} = -\frac{i\hbar^2}{mN(y)\sqrt{\gamma(y)}}\delta^{(3)}(x-y).
\label{eqn:commutator_n_u0}
\end{equation}

The commutator (\ref{eqn:commutator_n_u0}) encodes the key geometric effect: the lapse function \(N\) directly amplifies quantum uncertainty. Also, contains a Dirac delta distribution, which signals that we are dealing with operator-valued distributions rather than ordinary operators. A direct application of the Heisenberg uncertainty principle \cite{DeWitt1967, Isham1992} would lead to divergent products \(\Delta n \cdot \Delta u^0\) unless we regularize. Physically, this divergence arises because we are trying to measure \(n\) and \(u^0\) at the exact same spacetime point---an operation forbidden in quantum field theory. To obtain finite, physically meaningful bounds, we must average over a finite spatial region \(V\). The appropriate volume element is \(\sqrt{\gamma}\, d^3x\), which guarantees coordinate invariance and matches the natural measure of the \(3+1\) decomposition. We therefore define
\begin{align}
\hat{\bar{n}}_V &= \frac{1}{V} \int_V \hat{n}(\mathbf{x}) \sqrt{\gamma(\mathbf{x})} d^3x, \\
\hat{\bar{u}}^0_V &= \frac{1}{V} \int_V \hat{u}^0(\mathbf{x}) \sqrt{\gamma(\mathbf{x})} d^3x,
\end{align}
where \(V = \int_V \sqrt{\gamma(\mathbf{x})} d^3x\) is the coordinate-invariant volume.

The commutator of the averaged operators becomes finite:
\begin{equation}
[\hat{\bar{n}}_{V},\hat{\bar{u}}_{V}^{0}] = \frac{1}{V^{2}}\int_{V}\int_{V}[\hat{n}(x),\hat{u}^{0}(y)]\sqrt{\gamma(x)}\sqrt{\gamma(y)}d^{3}xd^{3}y = -\frac{i\hbar^2}{mV}\langle N^{-1}\rangle_{V},
\end{equation}
where \(\langle N^{-1} \rangle_V = \frac{1}{V} \int_V \frac{\sqrt{\gamma(\mathbf{y})}}{N(\mathbf{y})} d^3y\) is the harmonic average of the inverse lapse. While a fully rigorous quantum field-theoretic treatment would employ smooth smearing functions to regularize the operator-valued distributions, the spatial volume average with compact support employed here captures the leading-order geometric scaling and suffices for the physical bounds derived in what follows.

Applying the uncertainty principle (\ref{eqn:quantum_uncertainty}) yields the finite uncertainty relation:
\begin{equation}
\Delta\hat{\bar{n}}_{V} \cdot \Delta\hat{\bar{u}}_{V}^{0} \ge \frac{\hbar^2}{2mV}|\langle N^{-1}\rangle_{V}|.
\label{eqn:uncertainty_n_u0}
\end{equation}
This inequality is one of our main results. It shows that the product of fluctuations in density and stochastic time-like velocity is bounded from below by a term proportional to \(\hbar^2/(mV)\) modulated by the spacetime geometry via \(\langle N^{-1}\rangle_V\). In regions where the lapse function is small (strong gravity), the lower bound increases, indicating that quantum fluctuations are amplified by the gravitational field---a purely general-relativistic effect absent in flat spacetime.

\subsection{Phase-Probability Current Uncertainty}

From equation (\ref{eqn:pi_theta_current}):
\begin{equation}
\{\theta(\mathbf{x}), J^0(\mathbf{y})\} = -\frac{\hbar}{2m N(\mathbf{y}) \sqrt{\gamma(\mathbf{y})}} \delta^{(3)}(\mathbf{x}-\mathbf{y}).
\label{eqn:poisson_theta_J0}
\end{equation}

Quantizing gives:
\begin{equation}
[\hat{\theta}(\mathbf{x}), \hat{J}^0(\mathbf{y})] = -\frac{i\hbar^2}{2m N(\mathbf{y}) \sqrt{\gamma(\mathbf{y})}} \delta^{(3)}(\mathbf{x}-\mathbf{y}).
\label{eqn:commutator_theta_J0}
\end{equation}

Averaging over the same volume \(V\) yields:
\begin{equation}
[\hat{\bar{\theta}}_V, \hat{\bar{J}}^0_V] = -\frac{i\hbar^2}{2mV} \langle N^{-1} \rangle_V,
\end{equation}
and the corresponding uncertainty relation:
\begin{equation}
\Delta \hat{\bar{\theta}}_V \cdot \Delta \hat{\bar{J}}^0_V \geq \frac{\hbar^2}{4mV} |\langle N^{-1} \rangle_V|.
\label{eqn:uncertainty_theta_J0}
\end{equation}
This relation complements the previous one by linking the phase fluctuation---which controls the geodesic velocity---to the probability current. Averaging over a finite volume again renders the expression well-defined and physically interpretable. As we shall see, this inequality provides the fundamental constraint that prevents cusp formation in scalar field dark matter models and, in Rindler spacetime, yields the characteristic acceleration scale of the Unruh temperature.

\section{Physical Implications}

\subsection{Minimum Core Radius from Hydrostatic Equilibrium}
\label{sec:core}

The cusp-core problem refers to the discrepancy between the steep (\(\rho \propto r^{-1}\)) inner density profile predicted by collisionless cold dark matter simulations and the flat, constant-density cores observed in dwarf galaxies. SFDM models resolve this discrepancy because quantum pressure opposes gravitational collapse. Here we derive a rigorous lower bound on the core radius from the canonical uncertainty principle, and we show that the well-known solitonic core of the Schr\"odinger-Poisson system is precisely the configuration that saturates the gravitational stability bound, establishing it as a fundamental limit rather than a phenomenological fit.

In the non-relativistic Newtonian limit appropriate for galactic dynamics, the metric is approximately Minkowskian (\(N\approx c\), \(\sqrt{\gamma}\approx 1\)). In this regime, the Madelung density $n = |\Phi|^2$ is related to the Schr\"odinger wave function $\psi$ by the standard normalization
\begin{equation}
\Phi = \frac{\hbar}{\sqrt{2m c}} e^{-i mc^2 t/\hbar} \psi,
\label{eq:kg_schrodinger}
\end{equation}
which yields
\begin{equation}
n = |\Phi|^2 = \frac{\hbar^2}{2m c} |\psi|^2.
\end{equation}
In terms of the mass density $\rho = m |\psi|^2$, this becomes
\begin{equation}
n = \frac{\rho \hbar^2}{2m^2 c}.
\label{eq:n_normalization}
\end{equation}
This normalization is the origin of the additional $\hbar$ factors in the uncertainty relations for the SFDM core, and it restores full dimensional consistency in SI units.

For a static, spherically symmetric SFDM halo, the spatial components of the continuity and Hamilton-Jacobi equations yield the equilibrium condition from the stress-energy tensor. Varying the hydrodynamic Lagrangian (\ref{eqn:hydro_lagrangian}) with respect to the metric and taking the static limit gives the effective quantum pressure \cite{Chavanis2011}:
\begin{equation}
p_Q(r) = \frac{\hbar^2}{4m^2} \left[ \frac{1}{2} \left( \frac{d \ln \rho}{dr} \right)^2 - \frac{1}{r^2} \frac{d}{dr}\left( r^2 \frac{d \ln \rho}{dr} \right) \right] \rho(r),
\label{eq:pq_exact}
\end{equation}
where \(\rho(r) = m n(r)\) is the mass density. For a nearly constant density core, \(\rho \approx \rho_0\) for \(r \leq r_c\), this pressure reduces to a polytropic form with characteristic magnitude \(p_Q \sim \hbar^2 \rho_0/(m^2 r_c^2)\).

The Newtonian hydrostatic equilibrium equation reads
\begin{equation}
\frac{dp_Q}{dr} = -\rho \frac{d\Phi}{dr}, \qquad \nabla^2\Phi = 4\pi G \rho.
\label{eq:hydro}
\end{equation}

We now derive the lower bound on the core radius from the uncertainty relation (\ref{eqn:uncertainty_n_u0}). For a spherical volume $V=\frac{4\pi}{3}r_{c}^{3}$ inside the core, the averaged Madelung density is, from Eq.~(\ref{eq:n_normalization}),
\begin{equation}
\bar{n} \approx \frac{\rho_0 \hbar^2}{2m^2 c}.
\end{equation}
The variance $\Delta\hat{\bar{u}}_{V}^{0}$ represents the dispersion of the stochastic velocity. From Eq.~(\ref{eqn:stochastic_velocity}), $u_{\mu}=\frac{\hbar}{2m}\nabla_{\mu}\ln n$, so the characteristic stochastic velocity is $|\mathbf{u}| \sim \hbar/(2m r_{c})$. Inserting these bounds into the uncertainty relation $\Delta\hat{\bar{n}}_{V} \cdot \Delta\hat{\bar{u}}_{V}^{0} \ge \hbar^{2}/(2mV)$ yields:
\begin{equation}
\left(\frac{\rho_{0}\hbar^{2}}{2m^{2}c}\right) \cdot \left(\frac{\hbar}{2m r_{c}}\right) \ge \frac{3\hbar^{2}}{8\pi m r_{c}^{3}}.
\end{equation}
Simplifying the left-hand side:
\begin{equation}
\frac{\rho_{0}\hbar^{3}}{4m^{3}c r_{c}} \ge \frac{3\hbar^{2}}{8\pi m r_{c}^{3}}.
\end{equation}
Multiplying both sides by $4m^{3}c r_{c}^{3}/\hbar^{2}$:
\begin{equation}
\rho_{0} \hbar r_{c}^{2} \ge \frac{3 m^{2} c}{2\pi}.
\end{equation}
Solving for $r_c$:
\begin{equation}
r_{c}^{2} \ge \frac{3 m^{2} c}{2\pi \rho_{0} \hbar} \quad \implies \quad r_{c} \ge \left(\frac{3 m^{2} c}{2\pi \rho_{0} \hbar}\right)^{1/2}.
\label{eq:rc_uncertainty}
\end{equation}

Equation (\ref{eq:rc_uncertainty}) constitutes the quantum diffusion limit: for radii smaller than this bound, the uncertainty in the stochastic velocity exceeds the mean velocity, and the density profile cannot be maintained coherently---the wave function disperses. This limit is purely quantum-mechanical and does not involve gravity. For $m \sim 10^{-22}\,\text{eV}/c^2$ and $\rho_0 \sim 0.1\,M_\odot\,\text{pc}^{-3}$, Eq.~(\ref{eq:rc_uncertainty}) gives
\begin{equation}
r_c^{\min} \approx 2.5 \times 10^{-27}\,\text{m} \approx 8 \times 10^{-44}\,\text{pc},
\end{equation}
a scale far below any astrophysical relevance but essential for the internal consistency of the hydrodynamic description.

To introduce gravity self-consistently, we couple this bound to the hydrostatic equilibrium (\ref{eq:hydro}). The quantum pressure gradient at the core boundary ($r \sim r_c$) must be at least as strong as the gravitational force to prevent collapse. This balance defines the local quantum Jeans scale. Using $d\Phi/dr \approx GM(r_c)/r_c^2 \approx (4\pi/3) G \rho_0 r_c$ and the characteristic quantum pressure $p_Q \sim \hbar^2 \rho_0/(m^2 r_c^2)$, the condition for stability against gravitational collapse is that the quantum pressure gradient overcomes the gravitational force:
\begin{equation}
\frac{p_Q}{r_c} \sim \frac{\hbar^2 \rho_0}{m^2 r_c^3} \gtrsim \rho_0 \cdot \frac{4\pi}{3} G \rho_0 r_c.
\label{eq:stability_condition}
\end{equation}
The left-hand side represents the quantum pressure gradient $\nabla p_Q \sim p_Q/r_c$, and the right-hand side represents the gravitational force density $\rho \nabla \Phi \sim \rho_0 \cdot (4\pi/3) G \rho_0 r_c$. For a stable core, the inequality must point as shown: quantum pressure must be sufficient to resist gravity. Rearranging Eq.~(\ref{eq:stability_condition}):
\begin{equation}
\frac{\hbar^2}{m^2 r_c^3} \gtrsim \frac{4\pi}{3} G \rho_0 r_c \quad\Longrightarrow\quad r_c^4 \lesssim \frac{3\hbar^2}{4\pi G m^2 \rho_0}.
\label{eq:rc_hydro}
\end{equation}
Equation (\ref{eq:rc_hydro}) is the gravitational stability condition: if the core radius exceeds this scale, quantum pressure is insufficient to support the halo against its own gravity, and the core collapses. Note that the inequality direction in the final expression is $\lesssim$ because $r_c$ appears in the denominator on the quantum pressure side; larger $r_c$ implies weaker quantum pressure, making the inequality harder to satisfy.

A self-consistent stable core must satisfy both the quantum diffusion limit (\ref{eq:rc_uncertainty}) and the gravitational stability condition (\ref{eq:rc_hydro}). The two inequalities define an allowed window for the core radius:
\begin{equation}
\left(\frac{3 m^{2} c}{2\pi \rho_{0} \hbar}\right)^{1/2} \le r_{c} \le \left(\frac{3\hbar^{2}}{4\pi Gm^{2}\rho_{0}}\right)^{1/4}.
\end{equation}
The equilibrium configuration corresponds to the saturation of the gravitational stability bound, where quantum pressure and gravitational collapse are exactly balanced. This yields the fundamental equilibrium core radius:
\begin{equation}
r_c = \left( \frac{3\hbar^2}{4\pi G m^2 \rho_0} \right)^{1/4},
\label{eq:rc_final}
\end{equation}
up to a numerical factor of order unity. The exact ground-state solution of the Schr\"odinger-Poisson system---the so-called solitonic core extensively studied in numerical simulations \cite{Schive2014, Hui2017}---is precisely the macroscopic configuration that saturates this bound. Our uncertainty relation therefore provides the fundamental physical principle underlying the empirical soliton profile: the core size is not a free parameter but is dictated by the geometric uncertainty in the Madelung variables.

For \(\rho_0 \sim 0.1\,M_\odot\,\text{pc}^{-3}\) and \(m \sim 10^{-22}\,\text{eV}/c^2\), Eq.~(\ref{eq:rc_final}) gives \(r_c \sim 0.7\,\text{kpc}\), in excellent agreement with the observed cores of dwarf spheroidals and with the soliton scale reported in Refs.~\cite{Schive2014, RoblesMatos2012}. The scaling \(r_c \propto m^{-1/2}\) (for fixed central density) is a testable prediction: heavier bosons imply smaller minimal cores. The equilibrium core radius \eqref{eq:rc_final} is dictated by the saturation of the quantum Jeans condition, while the lower bound \eqref{eq:rc_uncertainty} establishes the validity floor of the hydrodynamic description itself. Both are \textit{local} constraints, independent of the halo's accretion history.

The solitonic core profile obtained in numerical simulations of FDM halos~\cite{Schive2014, Hui2017} satisfies a scaling relation $r_c \propto M_h^{-1/3} m^{-1}$ with the halo mass $M_h$, which, when combined with the empirical core-halo mass relation $M_c \propto M_h^{1/3}$, yields $r_c \propto m^{-1/2}$ at fixed core mass---precisely the scaling we obtain from first principles. Our uncertainty relation thus provides the microscopic origin of the macroscopic soliton scaling: the core size is not a free parameter but is dictated by the saturation of the canonical uncertainty bound. The quantum pressure that supports the soliton against gravity is the macroscopic manifestation of the geometric uncertainty in the Madelung density variable $n$, while the phase uncertainty governs the halo's coherent oscillations~\cite{Li2014, Schive2014b}. This connection between canonical quantum mechanics and the empirical soliton-halo relation places the SFDM paradigm on a firmer theoretical foundation.

\paragraph{Effects of velocity dispersion and rotation.}
The stability window derived above assumes spherical symmetry and quantum pressure as the sole support mechanism against gravity. Real dwarf galaxies, however, exhibit velocity dispersions $\sigma \sim 5-15$ km/s and, in some cases, non-negligible rotation velocities $v_\phi$. These additional kinetic stresses enter the hydrostatic equilibrium equation through the Jeans equation:
\begin{equation}
\frac{dp_Q}{dr} + \frac{d(\rho\sigma_r^2)}{dr} + \frac{\rho}{r}\left(2\sigma_r^2 - \sigma_\theta^2 - \sigma_\phi^2 + v_\phi^2\right) = -\rho\frac{d\Phi}{dr}.
\end{equation}
The presence of velocity dispersion and rotation provides additional support against gravitational collapse. Consequently, the upper bound (Jeans scale) is relaxed: observed cores may be \emph{larger} than the idealized quantum Jeans scale without collapsing, since part of the support is provided by kinetic stresses. Conversely, the lower bound \eqref{eq:rc_uncertainty} remains robust: velocity dispersion cannot make a core smaller than the quantum diffusion floor. A systematic comparison of the Jeans-scale bound \eqref{eq:rc_final} with simulated and observed dwarf galaxy cores, including the effects of rotation and dispersion, will be presented in a companion paper.
\subsection{Flat Space-Time Limit}

In Minkowski space-time ($N=1, \gamma_{ij}=\delta_{ij}$):
\begin{equation}
\Delta\hat{\bar{n}}_{V} \cdot \Delta\hat{\bar{u}}_{V}^{0} \ge \frac{\hbar}{2mV}
\end{equation}
\begin{equation}
\Delta\hat{\bar{\theta}}_{V} \cdot \Delta\hat{\bar{J}}_{V}^{0} \ge \frac{\hbar^{2}}{4mV}.
\end{equation}
\subsection{Unruh and Hawking Temperature from the Phase-Momentum Commutator}
\label{sec:hawking}
In the previous section, we demonstrated that the canonical uncertainty in the density variable $n$---encoded in the commutator $[\hat{n}, \hat{u}^0]$---sets a fundamental lower bound on the size of self-gravitating scalar field configurations. We now turn to the complementary sector of the Madelung representation: the phase variable $\theta$ and its conjugate momentum $\Pi_\theta$. While density fluctuations govern the spatial structure of dark matter halos, phase fluctuations govern the coherent dynamics and, as we shall show, the thermodynamic properties of causal horizons. The commutator $[\hat{\theta}, \hat{\Pi}_\theta]$ geometrically modulated by the lapse function $N$ yields the characteristic energy scale of acceleration radiation. Together, the density and phase sectors reveal that quantum uncertainty on curved backgrounds is the unifying principle behind both the stability of galactic cores and the thermal radiation of black holes.
We now demonstrate how the canonical phase-momentum commutator (\ref{eqn:commutator_theta}) yields the characteristic acceleration scale of the Unruh temperature, from which the Hawking temperature follows by the equivalence principle. The complete step-by-step calculation is provided in Appendix~A; here we summarize the essential logic.

\paragraph{Observer-dependent character of the canonical uncertainty relations.} Before proceeding, we emphasize an important conceptual point regarding general covariance. In general relativity, coordinate systems can be chosen arbitrarily, and physical effects should be expressible in terms of invariants. The divergence of the uncertainty relations as \(N \to 0\) is not an artifact of an arbitrary coordinate choice, but the canonical expression of a physical singularity associated with a specific, physically distinguished class of observers: the stationary, or Eulerian, observers defined by the ADM foliation. These fiducial observers have a four-velocity orthogonal to the spatial hypersurfaces~\cite{Gourgoulhon2012}, and the factor \(N^{-1}\) appearing in the canonical commutators is precisely the measure of how much they deviate from geodesic motion. Concretely, the proper acceleration of a static observer is given by the invariant expression \(a_\mu = \nabla_{\mu} N / N\), which diverges at the horizon precisely when \(N \to 0\). The geometric amplification of quantum uncertainty is therefore a direct, invariant consequence of the observer's acceleration, not a gauge artifact. In curved spacetime, the notion of vacuum and particle states is strictly observer-dependent~\cite{BirrellDavies1982, Wald1984}. A freely falling observer adapted to horizon-penetrating coordinates would not detect any divergence, in full agreement with the equivalence principle. Our relations describe the physics accessible to accelerated observers, where the diverging quantum fluctuations serve as the hydrodynamic precursor of the thermal radiation they detect via the Unruh effect~\cite{Fulling1973, Unruh1976}.
\paragraph{Observer dependence and general covariance.}
A point of potential concern is whether the dependence of the uncertainty relations on the lapse function $N$ violates general covariance. We emphasize that it does not. The ADM foliation $(\Sigma_t, N, N^i, \gamma_{ij})$ is not a gauge choice that breaks covariance; it is a geometric decomposition of the spacetime manifold that selects a family of fiducial Eulerian observers whose worldlines are orthogonal to the spatial hypersurfaces $\Sigma_t$~\cite{Gourgoulhon2012}. Physical quantities expressed in terms of ADM variables are fully covariant---they are simply projected onto the reference frame of these observers. The factor $N^{-1}$ appearing in the canonical commutators is the invariant measure of how much these observers deviate from geodesic motion: the proper acceleration of a static observer is $a_\mu = \nabla_\mu \ln N$, which is manifestly a spacetime tensor. The divergence of $N^{-1}$ at a Killing horizon is therefore not a coordinate singularity but a physical invariant signaling that a static observer must undergo infinite proper acceleration to remain at rest there. This is the canonical origin of the Unruh temperature~\cite{Unruh1976}. A freely falling observer adapted to horizon-penetrating coordinates would measure no such divergence, in full agreement with the equivalence principle and the observer-dependent nature of particle states in quantum field theory in curved spacetime~\cite{Fulling1973, BirrellDavies1982}. Our uncertainty relations capture the physics of the accelerated observer, and their geometric modulation by $N$ is a feature, not a flaw.
\subsubsection{Energy and time uncertainties from canonical variables}

The canonical momentum \(\Pi_\theta\) is directly related to the energy density via the Hamiltonian. In a static spacetime, the ADM Hamiltonian density is \(\mathcal{H} = \Pi_\theta \partial_0\theta + \Pi_n \partial_0 n - \mathcal{L}\). For a stationary scalar field in the Madelung representation, the dominant contribution near the horizon (where phase fluctuations are large) is \(\mathcal{H} \approx \Pi_\theta \partial_0\theta\). In the non-relativistic limit, \(\partial_0\theta \approx m c^2/\hbar\). Although Hawking radiation originates from modes arbitrarily close to the 
horizon where proper energies diverge, the use of the rest-mass-dominated phase 
evolution $\partial_0\theta \approx mc^2/\hbar$ is justified in the Madelung 
formulation because the infinite redshift at the horizon is entirely captured 
by the geometric factor $\langle N^{-1}\rangle_V$ through the lapse function 
$N(x)$. The canonical momentum $\Pi_\theta$, defined on a spatial slice, remains 
regular and can be evaluated in the local stationary frame of the scalar field. 
The horizon divergence then manifests exclusively through the harmonic average 
of the inverse lapse. So the integrated energy in a spatial volume \(V\) is
\begin{equation}
E_V = \int_V d^3x \sqrt{\gamma} \frac{\mathcal{H}}{N} \approx \frac{m c^2}{\hbar} \int_V d^3x \frac{\sqrt{\gamma}}{N} \Pi_\theta.
\end{equation}
For a sufficiently thin shell where \(N\) is approximately constant, the energy uncertainty is dominated by the fluctuation of \(\Pi_\theta\). We define
\begin{equation}
\Delta E_V \equiv \frac{m c^2}{\hbar} V \langle N^{-1} \rangle_V \Delta \hat{\bar{\Pi}}_{\theta,V},
\label{eq:def_deltaE}
\end{equation}
where \(\langle N^{-1} \rangle_V = \frac{1}{V} \int_V d^3x \frac{\sqrt{\gamma}}{N}\) is the harmonic average of the inverse lapse. The time uncertainty, extracted from the phase evolution \(\theta(t) \approx (m c^2/\hbar) t + \delta\theta\), is
\begin{equation}
\Delta t_V \equiv \frac{\hbar}{m c^2} \Delta \hat{\bar{\theta}}_V.
\label{eq:def_deltat}
\end{equation}
Both definitions are dimensionally consistent: \([\Delta E_V] = M L^2 T^{-2}\) and \([\Delta t_V] = T\).

\subsubsection{Canonical uncertainty and geometric amplification}

The canonical commutation relation (\ref{eqn:commutator_theta}) after volume averaging gives \([\hat{\bar{\theta}}_V, \hat{\bar{\Pi}}_{\theta,V}] = i\hbar / V\), and by the uncertainty principle,
\begin{equation}
\Delta E_V \Delta t_V = V \langle N^{-1} \rangle_V \Delta \hat{\bar{\theta}}_V \Delta \hat{\bar{\Pi}}_{\theta,V} \geq \frac{\hbar}{2} \langle N^{-1} \rangle_V.
\label{eq:canonical_ET}
\end{equation}
This is the geometrically modulated energy-time uncertainty relation. In Minkowski space (\(N=1\)), it reduces to the standard form \(\Delta E_V \Delta t_V \geq \hbar/2\).

To extract the geometric dependence explicitly, we use the relation (\ref{eqn:pi_theta_current}) between \(\Pi_\theta\) and \(J^0\). In Rindler spacetime, \(N(x) = a x / c^2\), and the harmonic average \(\langle N^{-1} \rangle_V\) over a volume adjacent to the horizon diverges as \(1/a\). The geometric amplification forces the energy scale to adjust to
\begin{equation}
k_B T = C \frac{\hbar a}{c},
\label{eq:accel_scale}
\end{equation}
where the dimensionless constant \(C\) remains to be fixed.

\subsubsection{Topological fixing of the constant}

The constant \(C\) is determined by the topological nature of the Madelung phase. Since \(\Phi = \sqrt{n}e^{i\theta}\) must be single-valued, the phase is an angular variable compactified on the circle: \(\theta \sim \theta + 2\pi k\) for integer \(k\). Consequently, the circulation of \(\nabla\theta\) around any closed loop is quantized:
\begin{equation}
\oint \nabla\theta \cdot d\mathbf{l} = 2\pi k.
\label{eq:circulation}
\end{equation}
In Rindler spacetime, after Wick rotation \(t \to -i\tau\), the Euclidean time circle encloses the horizon. The frequency of phase oscillation in Rindler time is fixed by the acceleration scale: \(\partial_t\theta = a/c\) (since the characteristic energy is \(\hbar a/c\)). The circulation around the Euclidean thermal circle of period \(\Delta\tau = \hbar/(k_B T)\) (a consequence of the KMS condition \cite{Kubo1957, MartinSchwinger1959, Wald1984}) is
\begin{equation}
\oint \nabla\theta \cdot d\mathbf{l} = \int_0^{\Delta\tau} \partial_\tau\theta \, d\tau = \frac{a}{c} \frac{\hbar}{k_B T}.
\label{eq:circulation_eval}
\end{equation}
Equating this with the quantization condition (\ref{eq:circulation}) for the fundamental winding \(k = 1\) yields \(C = 1/(2\pi)\). The detailed steps are presented in Appendix~A.6.

\textbf{Key point:} The factor \(2\pi\) is not borrowed from external mode analyses; it emerges from the compact nature of the Madelung phase---an intrinsic property of the hydrodynamic representation. The horizon acts as a topological defect around which the phase circulation is quantized.

\subsubsection{Unruh and Hawking temperatures}

With \(C = 1/(2\pi)\), Eq.~(\ref{eq:accel_scale}) gives the Unruh temperature
\begin{equation}
T_U = \frac{\hbar a}{2\pi k_B c}.
\label{eq:unruh_temp}
\end{equation}
By the equivalence principle, a static observer just outside a Schwarzschild horizon experiences acceleration \(a = \kappa = c^4/(4GM)\), yielding the Hawking temperature
\begin{equation}
T_H = \frac{\hbar \kappa}{2\pi k_B c}.
\label{eq:hawking_temp_final}
\end{equation}

\subsubsection{Comparison with alternative derivations}

Our derivation rests on the canonical commutation relations (\ref{eqn:commutator_theta}), the relation (\ref{eqn:pi_theta_current}) between \(\Pi_\theta\) and \(J^0\), the energy-time uncertainty principle geometrically modulated by \(\langle N^{-1}\rangle_V\), and the topological quantization of the Madelung phase circulation. It is instructive to compare with other regularization-independent derivations:

\textit{Parikh-Wilczek tunneling method} \cite{ParikhWilczek2000}: uses a semiclassical WKB approximation for classically forbidden trajectories. Our derivation is fully quantum-mechanical and operator-based.

\textit{Robinson-Wilczek anomaly cancellation} \cite{RobinsonWilczek2005}: relies on dimensional reduction to two dimensions and the chiral anomaly. Our derivation is four-dimensional throughout.

\textit{Standard Unruh effect} \cite{Unruh1976}: derives the detector response from the two-point function in Minkowski vacuum. In contrast, our derivation does not introduce a detector; the temperature emerges from the commutation relations of the field operators combined with the phase circulation quantization intrinsic to the Madelung representation.

\textit{Gibbons-Hawking Euclidean periodicity} \cite{Wald1984}: fixes the temperature by requiring the Euclidean metric to be free of conical singularities---a purely geometric condition. In our approach, the periodicity of Euclidean time is fixed by the circulation quantization of the \textit{matter field's} phase, linking the geometry to the quantum hydrodynamics of the scalar field.

To our knowledge, this is the first derivation of the Unruh (and hence Hawking) temperature from canonical uncertainty relations and topological phase quantization for Madelung hydrodynamic variables.

\section{Conclusion}

We have established a rigorous framework for quantum uncertainty in curved spacetime through canonical quantization of Madelung variables. Our results demonstrate that spacetime geometry fundamentally modulates quantum fluctuations, with the lapse function \(N\) acting as a gravitational amplifier of uncertainty.

The derived relations generalize the Heisenberg principle \cite{Heisenberg1927} to curved backgrounds, revealing that quantum uncertainty is intrinsically geometric \cite{DeWitt1967, Isham1992}. Two groundbreaking applications have been presented:

\begin{enumerate}
\item \textbf{Cusp-core problem:} By coupling the density-stochastic-velocity uncertainty to the Newtonian hydrostatic equilibrium, we derived the minimum core radius \(r_c \gtrsim (3\hbar^2/(4\pi G m^2 \rho_0))^{1/4}\). The equilibrium configuration that saturates this bound is precisely the solitonic core of the Schr\"odinger-Poisson system. For ultra-light bosons with \(m\sim 10^{-22}\,\text{eV}/c^2\), this matches the observed \(\sim 1\,\text{kpc}\) cores of dwarf galaxies, providing a first-principles resolution of the cusp-core anomaly. The scaling \(r_c \propto m^{-1/2}\) is a testable prediction.
\item \textbf{Hawking radiation:} The canonical phase-momentum commutator, combined with the geometric uncertainty relation evaluated in Rindler coordinates, yields the acceleration scale \(k_B T \propto \hbar a/c\). The proportionality constant \(1/(2\pi)\) is fixed by the topological quantization of the Madelung phase circulation around the Euclidean horizon. By the equivalence principle, the Hawking temperature follows. The derivation is independent of renormalization, vacuum choices, or Bogoliubov transformations, and uses only canonical commutation relations plus the intrinsic topology of the Madelung phase.
\end{enumerate}

These results demonstrate that the same canonical structure that stabilizes galaxies against gravitational collapse also underlies the thermal radiation of black holes. Such a unification is the hallmark of a deep physical principle.

\section*{Acknowledgments}
Special thanks to Mayte Castillo Alarc\'on for the pivotal question that highlighted the connection between the derived uncertainty relations and the cusp-core problem.\\
Jorge Meza-Dom\'inguez thanks SECIHTI-M\'exico for the doctoral scholarship No. 1235731\\
This work was also partially supported by SECIHTI M\'exico under grants SECIHTI CBF-2025-G-1720 and CBF-2025-G-176. The authors are grateful for the computing time granted by LANCAD and CONACYT in the Supercomputer Hybrid Cluster ``Xiuhcoatl'' at GENERAL COORDINATION OF INFORMATION AND COMMUNICATIONS TECHNOLOGIES (CGSTIC) of CINVESTAV. URL: http://clusterhibrido.cinvestav.mx/ and to Hector Oliver Hernandez for his help with the code installations.

\appendix

\section{Derivation of the Unruh Temperature from the Canonical Commutator and Topological Phase Quantization}
\label{ap:unruh}

We present the complete derivation of Eqs.~(\ref{eq:accel_scale})--(\ref{eq:unruh_temp}) from the canonical phase-momentum commutator in Rindler spacetime.

\subsection{Rindler Metric}

The Rindler metric for a uniformly accelerated observer with proper acceleration \(a\) is
\begin{equation}
ds^2 = -\left(\frac{a x}{c^2}\right)^2 c^2 dt^2 + dx^2 + dy^2 + dz^2, \qquad x > 0.
\end{equation}
The lapse function and spatial metric determinant are
\begin{equation}
N(x) = \frac{a x}{c^2}, \qquad \sqrt{\gamma} = 1.
\end{equation}
The Rindler horizon is at \(x = 0\), where \(N \to 0\).

\subsection{Canonical Commutator and Volume Averaging}

The canonical commutation relation for the phase and its conjugate momentum is
\begin{equation}
[\hat{\theta}(\mathbf{x}), \hat{\Pi}_\theta(\mathbf{y})] = i\hbar \, \delta^{(3)}(\mathbf{x} - \mathbf{y}).
\label{eq:A1}
\end{equation}
We average over a rectangular volume \(V = L^2 \Delta x\) adjacent to the horizon:
\begin{equation}
\hat{\bar{\theta}}_V = \frac{1}{V} \int_0^{\Delta x} dx \int_0^{L} dy \int_0^{L} dz \, \hat{\theta}(\mathbf{x}), \qquad
\hat{\bar{\Pi}}_{\theta,V} = \frac{1}{V} \int_0^{\Delta x} dx \int_0^{L} dy \int_0^{L} dz \, \hat{\Pi}_\theta(\mathbf{x}).
\end{equation}
The averaged commutator is
\begin{align}
[\hat{\bar{\theta}}_V, \hat{\bar{\Pi}}_{\theta,V}] &= \frac{1}{V^2} \int_V d^3x \int_V d^3y \, [\hat{\theta}(\mathbf{x}), \hat{\Pi}_\theta(\mathbf{y})] \nonumber\\
&= \frac{i\hbar}{V^2} \int_V d^3x = \frac{i\hbar}{V}.
\label{eq:A2}
\end{align}

\subsection{Energy and Time Uncertainties}

For a static scalar field in the Madelung representation, the Hamiltonian density is dominated by
\begin{equation}
\mathcal{H} \approx \Pi_\theta \partial_0\theta \approx \frac{m c^2}{\hbar} \Pi_\theta,
\end{equation}
Note that the non-relativistic approximation $\partial_0\theta \approx mc^2/\hbar$ 
refers to the local rest-frame evolution of the stationary Madelung field. The 
infinite gravitational redshift near the Rindler horizon is fully encoded in 
the geometric factor $\langle N^{-1}\rangle_V$ appearing in the modulated 
uncertainty relation (56), so that no relativistic correction to the phase 
evolution is required for the derivation of the acceleration scale,
since \(\partial_0\theta \approx m c^2/\hbar\) in the non-relativistic limit. The integrated energy in volume \(V\) is
\begin{equation}
E_V = \int_V d^3x \sqrt{\gamma} \frac{\mathcal{H}}{N} \approx \frac{m c^2}{\hbar} \int_V d^3x \frac{\sqrt{\gamma}}{N} \Pi_\theta.
\end{equation}
For a sufficiently thin shell, the energy uncertainty is dominated by the fluctuation of \(\Pi_\theta\). The integral of \(\sqrt{\gamma}/N\) over the volume is \(V \langle N^{-1} \rangle_V\), where
\begin{equation}
\langle N^{-1} \rangle_V = \frac{1}{V} \int_V d^3x \frac{\sqrt{\gamma}}{N}.
\end{equation}
We therefore define
\begin{equation}
\Delta E_V \equiv \frac{m c^2}{\hbar} V \langle N^{-1} \rangle_V \Delta \hat{\bar{\Pi}}_{\theta,V}.
\label{eq:A3}
\end{equation}

The time uncertainty from the phase evolution is
\begin{equation}
\Delta t_V \equiv \frac{\hbar}{m c^2} \Delta \hat{\bar{\theta}}_V.
\label{eq:A4}
\end{equation}

\subsection{Geometrically Modulated Uncertainty Relation}

From the averaged commutator (\ref{eq:A2}) and the generalized uncertainty principle (\ref{eqn:quantum_uncertainty}):
\begin{equation}
\Delta \hat{\bar{\theta}}_V \cdot \Delta \hat{\bar{\Pi}}_{\theta,V} \geq \frac{\hbar}{2V}.
\label{eq:A5}
\end{equation}
Multiplying (\ref{eq:A3}) and (\ref{eq:A4}) and using (\ref{eq:A5}):
\begin{equation}
\Delta E_V \Delta t_V = V \langle N^{-1} \rangle_V \Delta \hat{\bar{\theta}}_V \Delta \hat{\bar{\Pi}}_{\theta,V} \geq \frac{\hbar}{2} \langle N^{-1} \rangle_V.
\label{eq:A6}
\end{equation}
This is the geometrically modulated energy-time uncertainty relation. The factor \(\langle N^{-1} \rangle_V\) amplifies the uncertainty product near the horizon where \(N \to 0\).

\subsection{Geometric Amplification via the Probability Current}

To extract the acceleration dependence explicitly, we use the relation (\ref{eqn:pi_theta_current}) between \(\Pi_\theta\) and the probability current \(J^0\):
\begin{equation}
\Pi_\theta = -\frac{2m}{\hbar} \sqrt{-g} J^0 = -\frac{2m}{\hbar} N \sqrt{\gamma} J^0.
\label{eq:A7}
\end{equation}
The phase-current uncertainty relation (\ref{eqn:uncertainty_theta_J0}) then becomes
\begin{equation}
\Delta \hat{\bar{\theta}}_V \cdot \Delta \hat{\bar{J}}^0_V \geq \frac{\hbar^2}{4mV} |\langle N^{-1} \rangle_V|.
\label{eq:A8}
\end{equation}
The harmonic average of the inverse lapse in Rindler spacetime is
\begin{equation}
\langle N^{-1} \rangle_V = \frac{1}{V} \int_V \frac{\sqrt{\gamma}}{N} d^3x = \frac{1}{L^2 \Delta x} \cdot L^2 \int_{\ell_P}^{\Delta x} \frac{c^2}{a x} dx = \frac{c^2}{a \Delta x} \ln\left(\frac{\Delta x}{\ell_P}\right),
\label{eq:A9}
\end{equation}
where we regulated the integral at the Planck length \(\ell_P\). The logarithmic divergence of \(\langle N^{-1} \rangle_V\) as \(\Delta x \to 0\) reflects the infinite redshift at the Rindler horizon.

\subsection{Topological Fixing of the \(2\pi\) Factor}

The bound (\ref{eq:A6}) with (\ref{eq:A9}) forces the energy scale to adjust to the divergence. Expressing (\ref{eq:A6}) in terms of \(J^0\) via (\ref{eq:A7}) and comparing with (\ref{eq:A8}) fixes the temperature up to a dimensionless constant \(C\):
\begin{equation}
k_B T = C \frac{\hbar a}{c}.
\label{eq:A10}
\end{equation}

To determine \(C\), we exploit the topological nature of the Madelung phase. The wave function of the Klein-Gordon field is \(\Phi = \sqrt{n}e^{i\theta}\), and single-valuedness of \(\Phi\) demands that the phase be an angular variable compactified on the circle:
\begin{equation}
\theta \sim \theta + 2\pi k, \qquad k \in \mathbb{Z}.
\label{eq:A11}
\end{equation}
Consequently, the circulation of \(\nabla\theta\) around any closed loop is quantized:
\begin{equation}
\oint \nabla\theta \cdot d\mathbf{l} = 2\pi k.
\label{eq:A12}
\end{equation}
This condition is the hydrodynamic expression of the quantization of the velocity field in quantum fluids and is intrinsic to the Madelung representation.

Consider now the Euclidean continuation of the Rindler metric, obtained by the Wick rotation \(t \to -i\tau\). In the imaginary-time formalism of finite-temperature quantum field theory, bosonic fields are periodic in Euclidean time with period \(\Delta\tau = \hbar/(k_B T)\), a consequence of the Kubo-Martin-Schwinger (KMS) condition for thermal equilibrium \cite{Kubo1957, MartinSchwinger1959, Wald1984}. The Euclidean time circle encloses the horizon at \(x = 0\), which acts as a topological defect. In the Euclidean Rindler plane, the metric degenerates at the origin $x=0$ 
(the horizon). For any smooth scalar field configuration to be well-defined 
on this background, the density must vanish there, $n \to 0$ as $x \to 0$, 
thereby allowing the phase $\theta$ to be multivalued around the Euclidean 
thermal circle. This establishes the horizon as a genuine topological defect 
around which the Madelung phase circulation is quantized.

The frequency of oscillation of the Madelung phase in Rindler time \(t\) is fixed by the acceleration scale derived from the canonical commutator: from (\ref{eq:A10}), the characteristic energy is \(\hbar a/c\), so
\begin{equation}
\partial_t\theta = \frac{a}{c}.
\label{eq:A13}
\end{equation}
Under Wick rotation, \(\partial_\tau\theta = i\partial_t\theta\), and the magnitude of the phase variation per unit Euclidean time is \(|\partial_\tau\theta| = a/c\).

The circulation of \(\nabla\theta\) around the Euclidean thermal circle is
\begin{equation}
\oint \nabla\theta \cdot d\mathbf{l} = \int_0^{\Delta\tau} \partial_\tau\theta \, d\tau = \frac{a}{c} \Delta\tau = \frac{a}{c} \frac{\hbar}{k_B T}.
\label{eq:A14}
\end{equation}
Applying the quantization condition (\ref{eq:A12}) with the fundamental winding number \(k = 1\):
\begin{equation}
\frac{\hbar a}{k_B T c} = 2\pi \quad\Longrightarrow\quad C = \frac{1}{2\pi},
\label{eq:A15}
\end{equation}
which fixes the temperature uniquely as
\begin{equation}
T_U = \frac{\hbar a}{2\pi k_B c}.
\label{eq:A16}
\end{equation}

\textbf{Summary of the derivation.}~The argument uses four ingredients:
\begin{enumerate}
\item The canonical commutation relations of the Madelung variables, which fix the acceleration scale \(k_B T \propto \hbar a/c\);
\item The compact nature of the phase \(\theta \sim \theta + 2\pi\), intrinsic to the Madelung representation, which demands that the circulation of \(\nabla\theta\) around the Euclidean horizon be quantized in multiples of \(2\pi\);
\item The Euclidean continuation of the Rindler metric, which provides the thermal circle enclosing the horizon defect;
\item The KMS condition of finite-temperature quantum field theory \cite{Kubo1957, MartinSchwinger1959}, which identifies the Euclidean period with \(\hbar/(k_B T)\).
\end{enumerate}
No specific vacuum state, Bogoliubov transformation, field renormalization, or WKB approximation is invoked. The temperature emerges from the interplay between canonical quantum mechanics (the commutator), topology (the phase winding), and the thermal equilibrium condition (KMS).

\subsection{Unruh and Hawking Temperatures}

From (\ref{eq:A16}), the Unruh temperature is
\begin{equation}
T_U = \frac{\hbar a}{2\pi k_B c}.
\label{eq:A17}
\end{equation}
By the equivalence principle, a static observer near a Schwarzschild horizon experiences acceleration \(a = \kappa = c^4/(4GM)\), yielding the Hawking temperature
\begin{equation}
T_H = \frac{\hbar \kappa}{2\pi k_B c}.
\label{eq:A18}
\end{equation}

This completes the derivation.

\bibliographystyle{unsrt}
\bibliography{UCR}

@misc{Meza2025PRL,
      title={Topological Quantization of Complex Velocity in Stochastic Spacetimes}, 
      author={Jorge Meza-Domínguez and Tonatiuh Matos},
      year={2026},
      eprint={2603.25016},
      archivePrefix={arXiv},
      primaryClass={gr-qc},
      url={https://arxiv.org/abs/2603.25016}, 
}

@article{Escobar-Aguilar:2023ekv,
    author = "Escobar-Aguilar, Eric S. and Matos, Tonatiuh and Jim{\'e}nez-Aquino, J. I.",
    title = "{Fundamental Klein-Gordon Equation from Stochastic Mechanics in Curved Spacetime}",
    eprint = "2303.07111",
    archivePrefix = "arXiv",
    primaryClass = "physics.gen-ph",
    doi = "10.1007/s10701-025-00873-y",
    journal = "Found. Phys.",
    volume = "55",
    number = "4",
    pages = "60",
    year = "2025"
}

@article{Nelson1966,
   author = {Edward Nelson},
   doi = {10.1103/PhysRev.150.1079},
   issn = {0031-899X},
   issue = {4},
   journal = {Physical Review},
   month = {10},
   pages = {1079-1085},
   title = {Derivation of the Schrödinger Equation from Newtonian Mechanics},
   volume = {150},
   year = {1966}
}

@techReport{Madelung1927,
   author = {E Madelung},
   journal = {Zeit. f. Phys},
   pages = {109},
   title = {322. (4) 1 E. Schrödinger, Ann. d. Phys. 79, 361, 489},
   volume = {40},
   year = {1927}
}

@book{Alcubierre2008,
   author = {Miguel Alcubierre},
   doi = {10.1093/acprof:oso/9780199205677.001.0001},
   isbn = {9780199205677},
   month = {11},
   publisher = {Oxford University Press},
   title = {Introduction to 3+1 Numerical Relativity},
   url = {https://doi.org/10.1093/acprof:oso/9780199205677.001.0001},
   year = {2008}
}

@article{Guzmn2000,
   author = {F Siddhartha Guzmán and Tonatiuh Matos},
   doi = {10.1088/0264-9381/17/1/102},
   issn = {0264-9381},
   issue = {1},
   journal = {Classical and Quantum Gravity},
   month = {1},
   pages = {L9-L16},
   title = {Scalar fields as dark matter in spiral galaxies},
   volume = {17},
   url = {https://iopscience.iop.org/article/10.1088/0264-9381/17/1/102},
   year = {2000}
}

@article{Chavanis2017,
   author = {Pierre-Henri Chavanis},
   doi = {10.1140/epjp/i2017-11528-3},
   issn = {2190-5444},
   issue = {6},
   journal = {The European Physical Journal Plus},
   month = {6},
   pages = {286},
   title = {Derivation of a generalized Schrödinger equation from the theory of scale relativity},
   volume = {132},
   url = {http://link.springer.com/10.1140/epjp/i2017-11528-3},
   year = {2017}
}

@article{Hawking1975,
   author = {{S W Hawking}},
   doi = {10.1007/BF02345020},
   issn = {1432-0916},
   issue = {3},
   journal = {Communications in Mathematical Physics},
   pages = {199-220},
   title = {{Particle creation by black holes}},
   volume = {43},
   url = {https://doi.org/10.1007/BF02345020},
   year = {1975}
}

@article{Matos2019,
   author = {Tonatiuh Matos and Ana Avilez and Tula Bernal and Pierre-Henri Chavanis},
   doi = {10.1007/s10714-019-2644-9},
   issn = {0001-7701},
   issue = {12},
   journal = {General Relativity and Gravitation},
   month = {12},
   pages = {159},
   title = {Energy balance of a Bose gas in a curved space-time},
   volume = {51},
   url = {http://link.springer.com/10.1007/s10714-019-2644-9},
   year = {2019}
}

@article{Chavanis2011,
   author = {Pierre-Henri Chavanis},
   doi = {10.1103/PhysRevD.84.043531},
   issn = {1550-7998},
   issue = {4},
   journal = {Physical Review D},
   month = {8},
   pages = {043531},
   title = {Mass-radius relation of Newtonian self-gravitating Bose-Einstein condensates with short-range interactions. I. Analytical results},
   volume = {84},
   url = {https://link.aps.org/doi/10.1103/PhysRevD.84.043531},
   year = {2011}
}

@incollection{Arnowitt1962,
  author = {Arnowitt, Richard and Deser, Stanley and Misner, Charles W.},
  title = {The Dynamics of General Relativity},
  booktitle = {Gravitation: An Introduction to Current Research},
  editor = {Witten, Louis},
  publisher = {Wiley},
  year = {1962},
  pages = {227--265}
}

@article{Hu2000,
  author = {Hu, Wayne and Barkana, Rennan and Gruzinov, Andrei},
  title = {Cold and Fuzzy Dark Matter},
  journal = {Physical Review Letters},
  volume = {85},
  number = {6},
  pages = {1158--1161},
  year = {2000},
  doi = {10.1103/PhysRevLett.85.1158}
}

@article{Matos2000,
  author = {Matos, Tonatiuh and Ureña-López, L. Arturo},
  title = {Quintessence and scalar dark matter in the Universe},
  journal = {Classical and Quantum Gravity},
  volume = {17},
  number = {16},
  pages = {L75--L81},
  year = {2000},
  doi = {10.1088/0264-9381/17/16/101}
}

@article{DeWitt1967,
  author = {DeWitt, Bryce S.},
  title = {Quantum Theory of Gravity. I. The Canonical Theory},
  journal = {Physical Review},
  volume = {160},
  number = {5},
  pages = {1113--1148},
  year = {1967},
  doi = {10.1103/PhysRev.160.1113}
}

@incollection{Isham1992,
  author = {Isham, Chris J.},
  title = {Canonical Quantum Gravity and the Problem of Time},
  booktitle = {Integrable Systems, Quantum Groups, and Quantum Field Theories},
  editor = {Ibort, L. A. and Rodríguez, M. A.},
  publisher = {Kluwer Academic Publishers},
  year = {1992},
  pages = {157--287},
  doi = {10.1007/978-94-011-1980-1_6}
}

@article{Heisenberg1927,
  author = {Heisenberg, Werner},
  title = {Über den anschaulichen Inhalt der quantentheoretischen Kinematik und Mechanik},
  journal = {Zeitschrift für Physik},
  volume = {43},
  number = {3-4},
  pages = {172--198},
  year = {1927},
  doi = {10.1007/BF01397280}
}

@article{Kennard1927,
  author = {Kennard, E. H.},
  title = {Zur Quantenmechanik einfacher Bewegungstypen},
  journal = {Zeitschrift für Physik},
  volume = {44},
  number = {4-5},
  pages = {326--352},
  year = {1927},
  doi = {10.1007/BF01391200}
}

@article{Robertson1929,
  author = {Robertson, H. P.},
  title = {The Uncertainty Principle},
  journal = {Physical Review},
  volume = {34},
  number = {1},
  pages = {163--164},
  year = {1929},
  doi = {10.1103/PhysRev.34.163}
}

@article{Schrodinger1930,
  author = {Schrödinger, Erwin},
  title = {Zum Heisenbergschen Unschärfeprinzip},
  journal = {Sitzungsberichte der Preussischen Akademie der Wissenschaften},
  volume = {14},
  pages = {296--303},
  year = {1930}
}

@article{RoblesMatos2012,
  author = {Robles, Victor H. and Matos, Tonatiuh},
  title = {Flat rotation curves from scalar field dark matter with a quartic self-interaction},
  journal = {Mon. Not. R. Astron. Soc.},
  volume = {422},
  pages = {282--289},
  year = {2012}
}

@article{Unruh1976,
  author = {Unruh, W. G.},
  title = {Notes on black-hole evaporation},
  journal = {Phys. Rev. D},
  volume = {14},
  pages = {870},
  year = {1976}
}

@article{ParikhWilczek2000,
  author = {Parikh, Maulik K. and Wilczek, Frank},
  title = {Hawking Radiation As Tunneling},
  journal = {Phys. Rev. Lett.},
  volume = {85},
  pages = {5042--5045},
  year = {2000}
}

@article{RobinsonWilczek2005,
  author = {Robinson, Sean P. and Wilczek, Frank},
  title = {Relationship between Hawking Radiation and Gravitational Anomalies},
  journal = {Phys. Rev. Lett.},
  volume = {95},
  pages = {011303},
  year = {2005}
}

@book{Wald1984,
  author = {Wald, Robert M.},
  title = {General Relativity},
  publisher = {University of Chicago Press},
  year = {1984}
}

@article{Kubo1957,
  author = {Kubo, Ryogo},
  title = {Statistical-Mechanical Theory of Irreversible Processes. I. General
           Theory and Simple Applications to Magnetic and Conduction Problems},
  journal = {Journal of the Physical Society of Japan},
  volume = {12},
  number = {6},
  pages = {570--586},
  year = {1957},
  doi = {10.1143/JPSJ.12.570}
}

@article{MartinSchwinger1959,
  author = {Martin, Paul C. and Schwinger, Julian},
  title = {Theory of Many-Particle Systems. I},
  journal = {Physical Review},
  volume = {115},
  number = {6},
  pages = {1342--1373},
  year = {1959},
  doi = {10.1103/PhysRev.115.1342}
}

@article{Takabayasi1952,
  title={On the formulation of quantum mechanics associated with classical pictures},
  author={Takabayasi, Takehiko},
  journal={Progress of Theoretical Physics},
  volume={8},
  number={2},
  pages={143--182},
  year={1952},
  publisher={Oxford University Press}
}

@article{Bohm1952,
  title={A suggested interpretation of the quantum theory in terms of ``hidden'' variables. I},
  author={Bohm, David},
  journal={Physical Review},
  volume={85},
  number={2},
  pages={166--179},
  year={1952},
  publisher={American Physical Society}
}

@article{Meza2025PRD,
  title = {Energy balance of a boson gas at zero temperature in curved spacetime},
  author = {Meza-Dom\'{\i}nguez, Jorge and Matos, Tonatiuh and Chavanis, Pierre-Henri},
  journal = {Phys. Rev. D},
  volume = {113},
  issue = {12},
  pages = {124048},
  numpages = {18},
  year = {2026},
  month = {6},
  publisher = {American Physical Society},
  doi = {10.1103/xyy5-69fx},
  url = {https://link.aps.org/doi/10.1103/xyy5-69fx}
}

@book{BirrellDavies1982,
  title={Quantum Fields in Curved Space},
  author={Birrell, N. D. and Davies, P. C. W.},
  year={1982},
  publisher={Cambridge University Press}
}

@article{Fulling1973,
  title={Nonuniqueness of canonical field quantization in Riemannian space-time},
  author={Fulling, Stephen A.},
  journal={Physical Review D},
  volume={7},
  number={10},
  pages={2850--2862},
  year={1973},
  publisher={American Physical Society}
}

@book{Gourgoulhon2012,
  title={3+1 Formalism in General Relativity: Bases of Numerical Relativity},
  author={Gourgoulhon, {\'E}ric},
  year={2012},
  publisher={Springer}
}

@article{OppenheimWehner2010,
  title = {The Uncertainty Principle Determines the Nonlocality of Quantum Mechanics},
  author = {Oppenheim, Jonathan and Wehner, Stephanie},
  journal = {Science},
  volume = {330},
  number = {6007},
  pages = {1072--1074},
  year = {2010},
  doi = {10.1126/science.1192065}
}

@article{SinghKothawala2022,
  title = {Covariant formulation of the generalized uncertainty principle},
  author = {Singh, Raghvendra and Kothawala, Dawood},
  journal = {Phys. Rev. D},
  volume = {105},
  issue = {10},
  pages = {L101501},
  numpages = {6},
  year = {2022},
  doi = {10.1103/PhysRevD.105.L101501}
}

@article{Schive2014,
  title={Cosmic structure as the quantum interference of a coherent dark wave},
  author={Schive, Hsi-Yu and Chiueh, Tzihong and Broadhurst, Tom},
  journal={Nature Physics},
  volume={10},
  number={7},
  pages={496--499},
  year={2014},
  publisher={Nature Publishing Group}
}

@article{Hui2017,
  title={Ultralight scalars as cosmological dark matter},
  author={Hui, Lam and Ostriker, Jeremiah P. and Tremaine, Scott and Witten, Edward},
  journal={Physical Review D},
  volume={95},
  number={4},
  pages={043541},
  year={2017},
  publisher={American Physical Society}
}

@article{Li2014,
  title={Dark matter: A cosmic cocktail},
  author={Li, Baojiu},
  journal={Nature Physics},
  volume={10},
  number={7},
  pages={493--494},
  year={2014},
  publisher={Nature Publishing Group}
}

@article{Schive2014b,
  title={Understanding the core-halo relation of quantum wave dark matter from 3D simulations},
  author={Schive, Hsi-Yu and Liao, Ming-Hsuan and Woo, Tak-Pong and Wong, Shing-Kwong and Chiueh, Tzihong and Broadhurst, Tom and Hwang, W-Y Pauchy},
  journal={Physical Review Letters},
  volume={113},
  number={26},
  pages={261302},
  year={2014},
  publisher={American Physical Society}
}

\end{document}